\documentclass[a4paper,twoside]{article}
\def\simgt{{\raise-.5ex\hbox{$\buildrel>\over\sim$}}\ }

\usepackage[authoryear]{natbib}
\usepackage[small,bf]{caption}
\bibpunct{(}{)}{;}{a}{}{,}
\usepackage{graphicx}
\date{} 
\newcommand{\etal}{{et~al.\null}}
\newcommand{\eg}{{e.g.,}}

\newcommand{\ie}{{i.e.,}}
\newcommand{\oiii}{[O~{\footnotesize III}]}
\newcommand{\nii}{[N~{\footnotesize II}]}
\title{\large\bf\flushleft The Planetary Nebula Luminosity Function:  
Pieces of the Puzzle}
\author{\parbox{\textwidth}{\flushleft
\vspace{-0.5cm}
{\it R. Ciardullo}\\
\vspace{0.4cm}
{\small Department of Astronomy \& Astrophysics, Penn State University,
University Park, PA  16802 USA}\\
{\small Email: rbc@astro.psu.edu}}}
\begin{document}
\twocolumn[
\begin{changemargin}{.8cm}{.5cm}
\begin{minipage}{.9\textwidth}
\vspace{-1cm}
\maketitle
%
%
\small{\bf Abstract:}
Extragalactic surveys in the emission line of \oiii\ $\lambda 5007$ have 
provided us with the absolute line strengths of large, homogeneous sets of 
planetary nebulae.  These data have been used to address a host of problems, 
from the measurement of the extragalactic distance scale, to the study of 
stellar populations.  I review our current understanding of the \oiii\ 
planetary nebula luminosity function (PNLF), and discuss some of the physical 
processes that effect its structure.  I also describe the features of the 
H$\alpha$ PNLF, a function that, upon first glance, looks similar to the 
\oiii\ PNLF, but which includes a very different set of objects.  Finally, 
I discuss recent measurements of $\alpha$, the number of PNe found in a 
stellar population, normalized to that population's bolometric luminosity.  
I show that, contrary to expectations, the values of $\alpha$ found in 
actively star-forming spirals is essentially the same as those measured in 
late-type elliptical and lenticular systems.  I discuss how this result sheds 
light on the physics of the planetary nebula phenomenon.


\medskip
\medskip
\end{minipage}
\end{changemargin}
]
\small

\section{Introduction}
Photometric surveys for extragalactic planetary nebulae (PNe) are useful for a 
host of reasons:  they produce test particles for studies of galactic
kinematics and chemical evolution, they probe facets of stellar evolution
that are unobservable with conventional techniques, and they yield unique 
insights into the properties of Galactic PNe.  PN photometry even allows 
us to measure the extragalactic distance scale to a precision that few other 
methods can reach.   These uses come about because, unlike Galactic samples of 
PNe, the planetary nebulae of other galaxies are all at the 
same distance and all have the same amount of foreground reddening.  Thus, we 
have access to the objects' absolute line-strengths, and, in particular, 
the planetary nebula luminosity function (PNLF).

The history of the PNLF dates back to the early 1960's, when 
\citet{hw63} modeled the PNe of the Magellanic Clouds as a set of 
spherical gas clouds surrounding non-evolving central stars.  Under this
assumption, the expected Balmer-line flux from a PN should go as
\begin{equation}
F \propto N_H N_e V \propto R^{-3} \propto t^{-3} 
\end{equation}
which, when written in terms of magnitudes, yields
\begin{equation}
t \propto 10^{M/7.5} \propto e^{0.307 M}
\end{equation}
Since the number of objects observed between magnitudes 
$M$ and $M + dM$ is proportional to the time spent between those magnitudes,
the implied luminosity function of Magellanic Cloud PNe is then
\begin{equation}
N(M) \propto {dt \over dM} \propto e^{0.307 M}
\label{powerlaw_lf}
\end{equation}
Obviously, the above argument grossly oversimplifies the physics of the
planetary nebula phenomenon.  Most PNe are not spherically symmetric, and 
their early evolution is governed more by the hydrodynamics of a fast 
wind/superwind interaction than by simple photoionization and expansion
\citep{kwok82, bf02, schon07}.  As a result, the ionized mass of a young
PN should increase with time and produce a distribution of PN luminosities 
that is flatter than that implied by equation~(\ref{powerlaw_lf}).
Furthermore, as a PN evolves, its opacity to ionizing photons and its 
ionization structure may change as well, adding dips and wiggles to the 
expected form of the PNLF\null.  Yet for all this,
the \citet{jacoby80} observations of faint PNe in the Large Magellanic Cloud 
seemed to confirm the simple \citet{hw63} law.  Although the LMC observations 
were conducted at 5007~\AA, rather than H$\alpha$ (since PNe are easier to 
identify in the \oiii\ line), the agreement between the observations and the
simple theory suggested that much of the complex physics associated with 
planetary nebulae was not crucial to understanding their faint, late stages of 
evolution.

\begin{figure}
\begin{center}
\includegraphics[scale=0.388, angle=0]{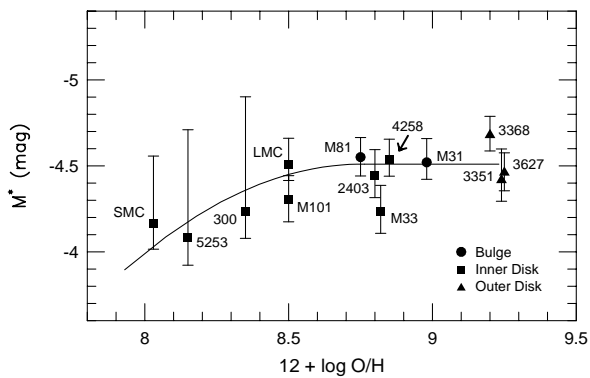}
\caption{Values of $M^*$ derived using the distances of 13 galaxies with
Cepheid measurements.  No correction for metallicity has been applied.
The abscissa shows the galaxies' oxygen abundances, as estimated from 
their H~II regions \citep{ferrarese}.   The curve shows the theoretical 
response of the nebula to metallicity \citet{djv92}, modified to reflect 
the presence of metal-poor stars in metal-rich galaxies.  The consistency 
of the measurements is excellent, and the scatter agrees with the internal 
errors of the methods.}
\end{center}
\label{pnlf_ceph}
\end{figure}

The late 1980's saw the introduction of the \oiii\ $\lambda 5007$ PNLF 
as an extragalactic standard candle.  Surveys of more than two dozen spiral 
and elliptical galaxies demonstrated that, while the faint-end of the PNLF 
may be fit by a \citet{hw63} exponential, the bright-end of the function 
truncates abruptly.  This led \citet{paper2} to propose fitting the PNLF 
using a cutoff exponential, 
\begin{equation}
N(M) \propto e^{0.307 M} \left\{ 1 - e^{3 (M^* - M)} \right\}
\label{eq_pnlf}
\end{equation} 
where $M^* = -4.48$ is the absolute luminosity of the brightest possible 
planetary in a magnitude system defined by
\begin{equation}
M_{5007} = -2.5 \log F_{5007} - 13.74 
\end{equation}
Remarkably, $M^*$ appeared to be virtually independent of stellar 
population: internal tests within galaxies \citep[\eg][]{chile, n5128}, 
internal tests within galaxy groups and clusters \citep[\eg][]{paper4, 
paper5, paper7}, and external tests in galaxies with distances from Cepheid 
and surface brightness fluctuation measurements \citep{paper12} yielded 
values of $M^*$ that were consistent to better than $\sim 10\%$.  As 
Figure~\ref{pnlf_ceph} illustrates, the only systematic trend was a
slight fading of $M^*$ at low metallicity -- a fading that was entirely
consistent with the expected response of a set of nebulae to lower 
oxygen abundance \citep{paper8, djv92, paper12}.

\section{The Faint-End Shape}
The constancy of $M^*$ means that the bright-end cutoff of the \oiii\ 
$\lambda 5007$ PNLF contains little or no information about the underlying 
stellar population.  However, the same is not necessarily true for the 
function's faint-end shape.  As described above, the physics of PN evolution
is extremely complicated, and a proper treatment of the problem requires
careful hydrodynamical modeling \citep[\eg][]{schon07}.  Yet at its most 
fundamental level, PN evolution is controlled by two timescales: one tied to
central star evolution, and the other to nebular expansion.  If most PNe 
are powered by low-mass, slowly-evolving central stars, then nebular expansion 
will dominate and the result will be a luminosity function that
monotonically increases towards fainter magnitudes.   Conversely, if the 
majority of central stars are high-mass objects, then their rapid stellar 
evolution will render nebular expansion unimportant.  In this case, the 
observed PNLF will primarily reflect the evolution of the central star's 
ionizing flux, and there will be a deficit of PNe with luminosities near the 
top of the white-dwarf cooling sequence.

\begin{figure}[t]
\begin{center}
\includegraphics[scale=0.318, angle=0]{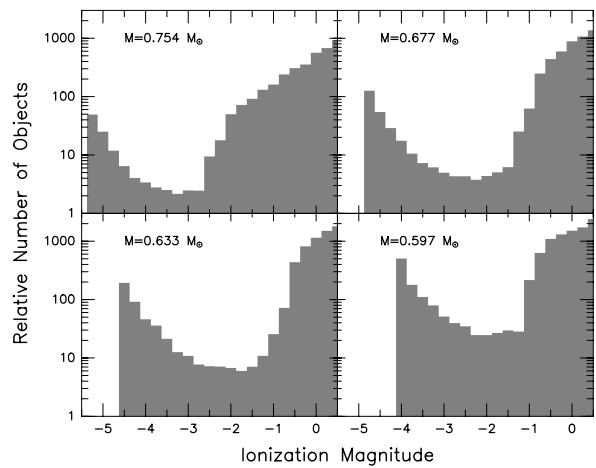}
\caption{The planetary nebula luminosity functions expected from nebulae
which reprocess their central star's ionizing radiation into \oiii\ 
$\lambda 5007$ flux at 10\% efficiency.   The histograms are based on
the post-AGB hydrogen-burning evolutionary tracks of \citet{vw94}.  Note the 
deficit of objects at intermediate magnitudes:  this is due to the rapid
cooling that occurs immediately after shell-burning ceases.  The figure
also suggests that for the highest-mass cores, the conversion efficiency 
of ionizing flux to \oiii\ $\lambda 5007$ radiation in no more than a 
few percent.}
\label{pagb}
\end{center}
\end{figure}

Figure~\ref{pagb} illustrates this effect by translating the post-AGB
evolutionary tracks of \citet{vw94} into a set of mock luminosity
functions, under the simplistic assumption that planetary nebulae
reprocess ionizing flux into \oiii\ $\lambda 5007$ radiation with 
10\% efficiency.  Obviously, many processes can reduce this
factor, and real PNe may evolve quite differently.  Nevertheless,
more sophisticated analyses bare out the result of this straightforward
analysis:  populations which are dominated by high core mass PNe 
must have a dip in their luminosity function \citep{mendez08}.

The stars of M31's bulge have ages of $\simgt 6$~Gyr
\citep{trager}.  Consequently, from the initial mass-final mass relation
\citep{kalirai}, we would expect the region to have PNe with low-mass,
slowly evolving central stars.  This is indeed the case:  
the deep \oiii\ luminosity function of \citet{paper12} shows that the
population's PNLF is well-fit by the \citet{hw63} law for non-evolving central
stars.  Conversely, the star-forming populations of the Small Magellanic
Cloud \citep{jd02} and M33 \citep{m33pn} should have many high-mass cores
that evolve rapidly across the HR diagram.  Again, this is what is seen,
as the luminosity functions of both systems have a clear
deficit of intermediate luminosity objects.  

\begin{figure}
\begin{center}
\includegraphics[scale=0.362, angle=0]{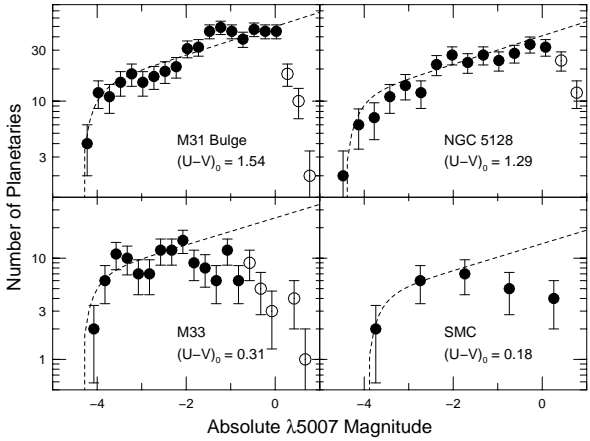}
\caption{Deep \oiii\ PNLFs for the bulge of M31, the envelope of
NGC~5128, the disk of M33, and the Small Magellanic Cloud.  Open circles
denote bins effected by incompleteness; the dotted line shows the
\citet{hw63} exponential law.   Only the strongly star-forming
systems have a deficit of PNe at intermediate magnitudes.}
\label{pnlfs}
\end{center}
\end{figure}

How fast is this transition?  In theory, we can constrain this quantity by 
examining the PNLF in a galaxy where star-formation has ceased in the recent 
past.  Unfortunately, such objects are difficult to identify.  One possible
location is the envelope of NGC~5128:  this interacting elliptical has
a blue color that is reminiscent of the inter-arm region of a spiral galaxy,
and still contains some evidence of recent star-formation \citep{rgz04}.
As Figure~\ref{pnlfs} illustrates, a deep planetary nebula survey of the 
region reveals a PNLF that is identical to that found for
M31's bulge; there is no hint of the mid-magnitude deficit present in the 
SMC and M33.  This suggests that the transition from a central-star 
driven luminosity function to a nebula-driven luminosity function must occur 
rather quickly:  unless the PN population is dominated by objects from very 
young progenitors, the \citet{hw63} law is the better predictor of the 
PNLF's shape.

\section{The PNLF in H$\alpha$}
Since the brightest PNe emit predominantly in \oiii\ $\lambda 5007$,
it is natural that PN surveys in distant galaxies focus on this wavelength.
But other emission lines contain information as well, and for many
Galactic surveys, it is the H$\alpha$ emission that is most-easily studied.  
Thus, we should consider the probative value of this brightest Balmer line
as well.

\begin{figure}[t]
\begin{center}
\includegraphics[scale=0.399, angle=0]{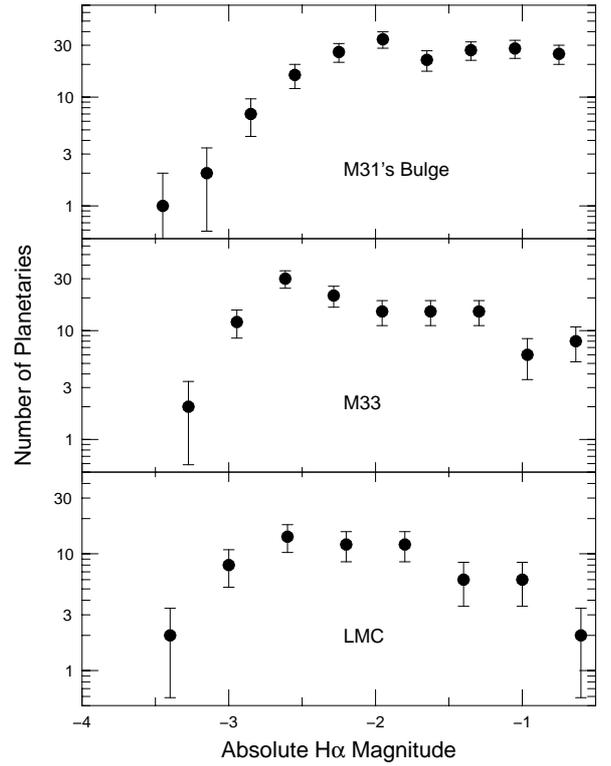}
\caption{The H$\alpha$ PNLFs for M31's bulge, M33's disk and the
LMC\null.  Though the shape of the cutoff differs from galaxy to galaxy, 
the brightest PN in each system has about the same absolute brightness.  
The turnover at faint magnitudes is partially due to incompleteness, as we 
are using \oiii\ selected samples.}
\label{ha_pnlfs}
\end{center}
\end{figure}

Figure~\ref{ha_pnlfs} displays the H$\alpha$ luminosity function for three
Local Group galaxies, using an H$\alpha$ magnitude defined via
\begin{equation}
M_{{\rm H}\alpha} = -2.5 \log F_{{\rm H}\alpha} - 13.67
\end{equation}
Like the \oiii\ PNLF, the H$\alpha$ luminosity function displays a cutoff
that is insensitive to stellar population.  The planetaries
produced by the old stellar population of M31's bulge have roughly
the same maximum H$\alpha$ flux as the PNe created in the star-forming 
environments of M33 and the Large Magellanic Cloud.  Although the cutoff
in the H$\alpha$ luminosity function is a factor of $\sim 2.5$ fainter than 
that for \oiii\  $\lambda 5007$, the data do suggest that H$\alpha$ 
measurements of PN ensembles can be used to estimate distance.

\begin{figure}[t]
\begin{center}
\includegraphics[scale=0.432, angle=0]{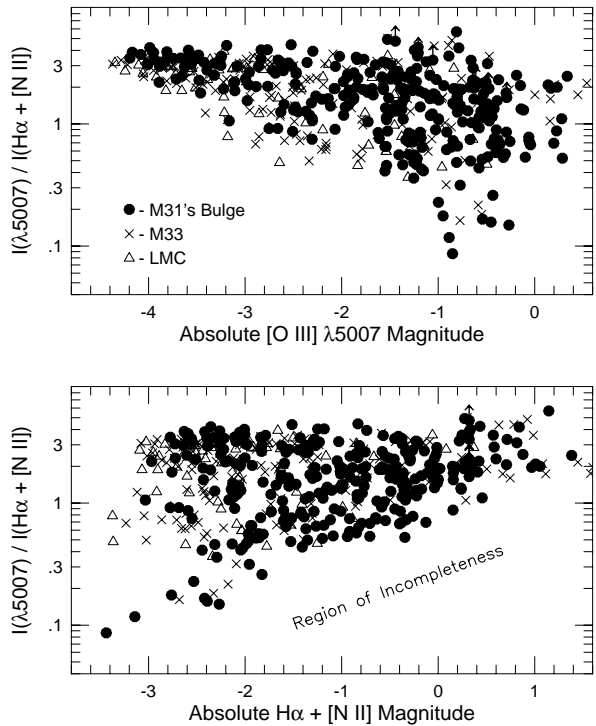}
\caption{The ratio of \oiii\ $\lambda 5007$ to H$\alpha$+\nii\ 
observed for PNe in M31's bulge, M33's disk, and the Large Magellanic
Cloud.  When plotted against \oiii\ brightness, the samples appear
rather homogeneous, especially at the bright end. When plotted 
against H$\alpha$ brightness, this consistency disappears.}
\label{squiggle2}
\end{center}
\end{figure}

This is not to say that all the PNe that are bright in H$\alpha$ are also
bright in \oiii.  In fact, when compared against the most luminous objects
in the \oiii\ $\lambda 5007$ PNLF, H$\alpha$-bright objects are surprisingly
inhomogeneous.  This can be seen in Figure~\ref{squiggle2},
which gives the \oiii\ to H$\alpha$+\nii\ line ratios of
PNe in M31's bulge, M33's disk, and throughout the Large Magellanic 
Cloud.  When plotted against \oiii\ absolute magnitude, the data show 
excellent consistency at the bright end, as the line-ratios, $R$, populate 
a wedge in emission-line space defined by
\begin{equation}
4 > R > 3.14 \left( {L \over L^*} \right)^{0.92} 
\end{equation}
where $L^* = 2.4 \times 10^{36}$~ergs~s$^{-1}$ in the monochromatic 
5007~\AA\ line \citep{herrmann}.  All PNe in the top magnitude of the 
luminosity function have \oiii\ brighter than H$\alpha$ by a considerable 
amount.  Conversely, PNe in the top magnitude of the H$\alpha$ PNLF have line
ratios that span the entire range observed, from $0.1 < R < 4$.  This
scatter, which has been reproduced in the hydrodynamic simulations of 
\citet{mendez08}, clearly demonstrates that more than one type of object 
is contributing to the bright-end of this function.

Figure~\ref{squiggle2} also illustrates a danger about drawing conclusions
from samples of objects selected via a single emission-line.  PNe in distant
galaxies are typically identified from their flux at \oiii\ $\lambda 5007$;
consequently, objects that are faint in \oiii\ but bright in H$\alpha$
may not be detected.   The result can be a censored dataset.
So, while the top panel of Figure~\ref{squiggle2} is an 
accurate representation of the distribution of objects in emission-line space, 
the bottom panel of the figure is seriously incomplete at the faint
end.   The same is true for the luminosity functions of Figure~\ref{ha_pnlfs}: 
although the deficits of PNe at faint magnitudes may be real, sample 
incompleteness is also playing an important role.  In the Milky
Way, where large samples of PNe are selected via their brightness in
H$\alpha$ \citep{mash, iphas}, this bias will be seen when objects are 
plotted against their \oiii\ $\lambda 5007$ magnitude.

\section{The PNLF and Distances in the Milky Way}
Figures~\ref{pnlfs}, \ref{ha_pnlfs}, and \ref{squiggle2} suggest a method of
constraining the distances to PNe within the Milky Way.  As the figures
illustrate, PNe populate a specific region of emission-line space:
there is a hard upper limit to a PN's luminosity, both in \oiii\ 
$\lambda 5007$ and H$\alpha$.  Moreover, objects near the bright-end
cutoff of the \oiii\ luminosity function are all high-excitation objects.  
Thus the line strengths of a Galactic planetary nebula immediately provide an
upper limit to distance.  For example, a PN
which is twice as bright in H$\alpha$ than in \oiii\ cannot be
any brighter than $M_{5007} \sim -2$.

\begin{figure}[t]
\begin{center}
\includegraphics[scale=0.432, angle=0]{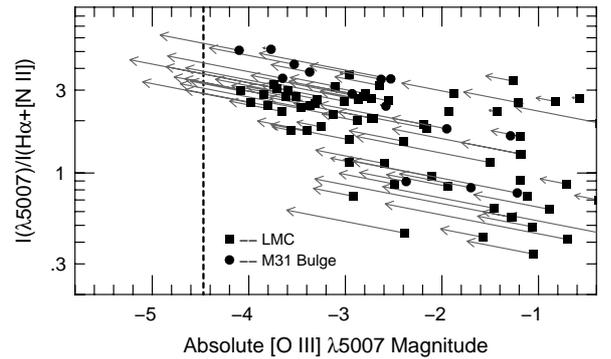}
\caption{The observed ratio of [O~III] $\lambda 5007$ to H$\alpha$+\nii\ 
plotted against [O~III] $\lambda 5007$ absolute magnitude for those bright PNe
in the LMC and M31 with published spectrophotometry.  The arrows display
the effects of internal extinction, as estimated from the objects' Balmer
decrement.  The location of the PNLF cutoff, $M^*$, is identified via the
dotted line.  Note that, although no PN is ever {\it observed\/} to have
an [O~III] $\lambda 5007$ magnitude brighter than $M^*$, many have intrinsic
luminosities much brighter than this value.}
\label{dust}
\end{center}
\end{figure}

Unfortunately, such relations are of limited use.   Like all objects,
PNe are affected by extinction along the line-of-sight.  This extinction
has two components: one from the foreground dust of the Milky Way
Galaxy, and the other from circumstellar material.  The effects of the former
have to be removed, since the dust is totally unrelated to the planetary.  
Conversely, the latter extinction must remain untouched, as it is an integral 
part of the planetary nebula phenomenon.  Indeed, as Figure~\ref{dust} 
illustrates, circumstellar extinction helps define the PNLF cutoff:  without 
this dust, many of the PNe observed in the Large Magellanic Cloud and M31 would 
be substantially brighter than $M^*$.  Thus, if one wants to use PN line 
strengths as a distance indicator, one has to somehow separate the Galactic 
and circumstellar components of extinction.

\section{The PNLF's Normalization}

An under-appreciated component of the PNLF is its normalization.  According
to the theory of stellar energy generation, all non-star forming stellar
populations should create 
$B \sim 2 \times 10^{-11}$~PNe~yr$^{-1}~L_{\odot}^{-1}$, regardless of
the system's age, metallicity, or initial-mass function \citep{renzini}.
Therefore, if one has an estimate for the mean PN lifetime, that age
can immediately be translated into an expected number of observable 
planetaries.  Any departure from this number provides a constraint on the
timescale of the PN phase and/or the importance of alternative 
channels of stellar evolution.

To perform this experiment, one needs to estimate $\alpha$, the number of PNe 
divided by the total bolometric luminosity of the underlying stellar
population.  This presents two problems.  First, one has to translate broadband
measurements of a galaxy's flux into total bolometric luminosity.  This 
is not as difficult as it sounds:  the PNLF provides the galaxy's distance,
and the population's bolometric correction can be assumed to be 
B.C. $\sim -0.85$, independent of its age or metallicity \citep{buzzoni}.  
Second, one can never observe the entire PN population of a distant galaxy:  
typically, one only counts objects in the top $\sim 1$~mag of the luminosity 
function.  Thus, extragalactic PN surveys usually quote 
$\alpha_{0.5}$ or $\alpha_{2.5}$, \ie\ the normalized 
number of PNe in the top 0.5 or 2.5~mag of the luminosity function.  To obtain 
the total PN population, this value has to be extrapolated over a range of at 
least 5 or 6 magnitudes.

For example, in their 1989 survey of the Leo~I Cloud, \citet{paper4} found 
values of $\alpha_{2.5} \sim 40 \times 10^{-9}$~PNe~$L_{\odot}^{-1}$ for
three different galaxies.  They then extrapolated this value to $\alpha_{8.0}$ 
using the \citet{hw63} law, and assumed a timescale for the PN phenomenon of 
$\sim 25,000$~yr \citet{pottasch84}.  The result was a stellar death rate 
that was surprisingly close to the theoretical value for single stars.

\begin{figure}
\begin{center}
\includegraphics[scale=0.362, angle=0]{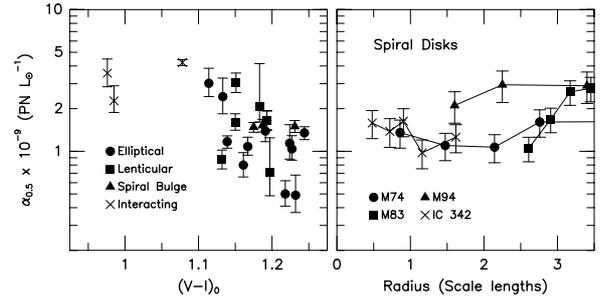}
\caption{The left hand panel displays values of $\alpha_{0.5}$ 
for late-type stellar populations; the right hand panel plots $\alpha$
measurements performed in the inter-arm regions of spiral disks.
Although there are trends in the data --- large, red ellipticals are less 
efficient at producing \oiii-bright PNe than smaller, blue galaxies --- the 
most striking feature is the similarity in the numbers.  All stellar 
populations have about 1 bright PN for every $5 \times 10^8$ solar 
(bolometric) luminosities.}
\label{alpha}
\end{center}
\end{figure}

Unfortunately, the true situation is a bit more complicated.
\citet{bs-pn} have presented values of $\alpha_{0.5}$ for $\sim 20$ elliptical
and lenticular galaxies.  As the left-hand panel of Figure~\ref{alpha} 
demonstrates, early-type systems typically have one bright PN (\ie\ within
0.5~mag of $M^*$) for every $\sim 5 \times 10^8 L_{\odot}$ of bolometric
light, with redder galaxies trending towards lower values of $\alpha$, and
bluer systems approaching the theoretical limit mentioned above.  But
remarkably, recent measurements of $\alpha_{0.5}$ in the disks of late-type
spirals have produced values of $\alpha$ in this same range, between 
1 and $3 \times 10^{-9}$~PNe$~L_{\odot}^{-1}$ \citep{herrmann}.   At 
first glance, the agreement between these numbers seems extraordinary.

There are several factors which must be considered before interpreting
this result.  The first, of course, is that much of a spiral galaxy's
luminosity comes from an extremely young stellar population that does not
produce planetary nebulae.  Specifically, if a galaxy has been producing 
stars at constant rate over a Hubble time, then, according to the models 
of \citet{bruzual}, roughly half of its bolometric light should come from 
populations formed less than $\sim 10^8$~years ago.   The luminosity from
these young stars should dilute the galaxy's PN-to-light ratio,
and cause our measurement of $\alpha$ to be lower than
what would otherwise be predicted.

On the other hand, estimates of the total PN population require an
extrapolation.  For simplicity, this is usually performed by integrating a
a fit to equation~(\ref{eq_pnlf}).  However, as Figure~\ref{pnlfs} shows,
star-forming systems have a deficit of objects at intermediate
magnitudes; if one integrates the SMC's PNLF found by \citet{jd02},
then one obtains a total PN population that is $\sim 1.6$ times less than
that derived from the exponential law.   Observations of the brightest
planetaries in late-type systems will therefore overestimate the total
number of PNe present in the galaxy.   Fortuitously, this error
is of the same order of magnitude as the luminosity dilution
and works in the opposite direction.   Thus, the observe values of 
$\alpha$ should probably not deviate very strongly from their nominal value.

Interpreting the values of $\alpha$ seen in elliptical galaxies is harder.
PNe at the bright-end cutoff of the \oiii\ PNLF emit $\sim 600 L_{\odot}$
monochromatically at 5007~\AA, and according to both models and observations,
no more than $\sim 10\%$ of the central star's flux can come out in this
line \citep[\eg][]{paper1, marigo, schon07}.  Thus, the central
stars of \oiii-bright PNe must be more luminous than $\sim 6000 L_{\odot}$;
in fact, in M31's bulge, 3 out of the 12 PNe analyzed by \citet{jc99} 
have central star luminosities in excess of $14,000 L_{\odot}$.
This is a problem, since in order to generate $\sim 6000 L_{\odot}$, a central 
star has to have a core mass of at least $0.6 M_{\odot}$ \citep{vw94}.  Cores
as massive as this require high-mass (young) progenitors \citep{kalirai},
and such objects do not exist in elliptical and lenticular systems.

There are two ways around this problem.  The first is to hypothesize that
objects at the bright-end of the \oiii\ PNLF are not PNe at all, but rather
some sort of symbiotic star \citep{soker}.  Such a solution solves the
luminosity problem at the expense of continuity:  why should such objects
have the same maximum luminosity and same luminosity function normalization 
as normal planetary nebulae?  The alternative is to invoke a new
channel of stellar evolution involving binary stars.  Binary interactions on 
the red giant or asymptotic giant branches are unlikely to be the answer,
since these encounters tend to truncate the stellar evolution process
before the creation of a massive core \citep{ibenlivio}.  However,
encounters on the main sequence, either through conservative mass-transfer 
between primordial close binaries \citep{mccrea} or through the dynamical 
evolution of triple systems \citep{perets} can create field blue straggler 
stars, and, as \citet{bs-pn} have pointed out, blue stragglers have the proper 
number, core properties, and evolutionary timescales to evolve into 
\oiii-bright planetaries.  In fact, if this scenario is correct, and the 
\citet{kozai} cycle-tidal friction process is responsible for creating blue 
straggler stars, then many of the brightest PNe we observe should have 
wide-binary companions \citep{perets}.

How important is the contribution of blue straggler descendents to the PNLFs 
of elliptical galaxies?  This is difficult to calculate, due to 
uncertainties in the initial mass-final mass relation at the low-mass end.
However, if we assume that old stellar populations are capable of
producing $\sim 0.55 M_{\odot}$ cores, and that these cores can, at 
best, convert ionizing radiation into \oiii\ $\lambda 5007$ flux with
10\% efficiency, then PNe from single stars can, theoretically, reach
$M_{5007} \sim -3$.  If the \oiii\ PNLF extends $\sim 8$~mag, then this 
implies that the brightest $\sim 5\%$ of planetary nebulae must come from 
an alternative channel of evolution.

\section{Prospects and Difficulties}
Over the years, attempts to model the PNLF have grown more and
more sophisticated.  Simple static models based on arbitrary distributions 
of central star masses \citep{paper1} have been replaced by simulations
which use realistic galactic star-formation histories \citep{marigo} and 
hydrodynamic star-nebular interactions \citep{mendez08}.  These analyses
have had some success.  By including the effects of metallicity 
on the nebula, \citet{djv92} were able to reproduce the fading of the PNLF 
cutoff in metal-poor populations \citep{paper12}.   Similarly, the 
hydrodynamical models of \citet{mendez08} have been able to accurately mimic 
the shape of the PNLF in young stellar systems, as well as the distribution of
PNe in \oiii\ $\lambda 5007$-H$\alpha$+[N~II] emission line space.

The trouble comes when trying to model the behavior of the PNLF with age.
\citet{djv92} and \citet{mendez08} both limited their analyses
to the actively star-forming populations of the Magellanic Clouds, with
good reason.  When considering single-star evolution, no reasonable 
initial mass-final mass relation can create the high mass cores needed 
to power $M^*$ planetaries.  This is illustrated vividly by the
simulations of \citet{marigo}:  if the PNLF were populated solely
by the remnants of single-stars, $M^*$ would fade by
$\sim 3$~mag in the first $\sim 6$~Gyr after the cessation of star
formation.  Clearly in order to reproduce the observed PNLF of old stellar
systems, another channel is needed to create PNe with relatively high
mass ($M > 0.6 M_{\odot}$) cores.

Another problem with current PNLF modeling efforts is that none 
include the effects of circumstellar extinction.   Most PNe are surrounded
by a circumstellar envelope of neutral gas and dust,
and, as Figure~\ref{dust} illustrates, the extinction associated with this
dust is an important contributor to the PNLF's cutoff.   In star-forming
systems, a typical \oiii\ bright planetary will be extinguished by 
$A_{5007} \sim 0.7$~mag \citep{kim2}, and reddening values two to three times
this number are not uncommon \citep[\eg][]{md1, md2}.   A similar range
of extinctions have been derived for the PNe of M31's bulge and M32
\citep{richer, jc99}, thus demonstrating that even in old systems, the effect
of dust on the PNLF cutoff cannot be neglected.  Of course, because extinction
estimates for statistically complete sets of extragalactic PNe are 
difficult to obtain, fully de-reddened PNLFs are not yet available to confront
the models.  Fortunately, this will soon change with the publication
of the results from the MASH survey of the LMC \citep{mash1}.

Finally, the PNLF models to date have all ignored the information 
contained in the PNLF's normalization.   A full description of the 
PNLF must not only explain its shape, but also the lower values of $\alpha$
observed in large, metal-rich elliptical galaxies, and the similarity
between the $\alpha$ values of spheroid and disk populations.  This may 
turn out to be the most difficult challenge of all, if, as currently
suggested, there is more than one scenario for planetary nebula formation.

\section{Conclusion}
Most of the data on extragalactic PNe comes from observations in the \oiii\ 
emission line at 5007~\AA.  But the \oiii\ PNLF is just a 
single one-dimensional projection of a more complicated function that 
distributes objects throughout a multi-dimensional phase-space that 
includes, for instance, H$\alpha$, H$\beta$, continuum 
emission at 4.5, 5.8, and 8.0~$\mu$m, and the emission-line strengths of 
nitrogen, sulfur, neon, argon, and helium.  Figure~\ref{ha_pnlfs} displays a
different one-dimension projection onto an axis which plots H$\alpha$, while
Figure~\ref{squiggle2} shows a pair of two-dimensional projections onto
planes which include the ratio of \oiii\ $\lambda 5007$ to H$\alpha$.
A fully successful model of the PNLF needs to not only reproduce these
projections, but also make predictions about the distribution of PNe
in other phase-space dimensions.  Similarly, the large homogeneous
surveys now being performed in the Galaxy and the Magellanic Clouds
have the potential to populate this PN phase space diagram as never before.
The resulting confrontation between theory and observations is sure
to generate new insights into the planetary nebula phenomenon.

\section*{Acknowledgments} 
We would like to thank the organizers of the MASH workshop for supporting this
review, and the anonymous referee for excellent comments on the first
draft of this text.  The work was supported by NSF grant AST 06-07416.


\end{document}